\documentstyle[11pt,fleqn]{article}

\makeatletter

\def\@revmess#1#2{\typeout{CJPhys LaTeX #1: #2}}
\ifx\selectfont\undefined %
\@revmess{message}{NFSS not detected. Assuming OFSS.}
\let\reset@font\relax
\def\mathhexbox{\protect\mathhexbox@}
\def\mathhexbox@#1#2#3{\relax
\ifmmode\mathpalette{}{\m@th\mathchar"#1#2#3}%
\else\leavevmode\hbox{$\m@th\mathchar"#1#2#3$}\fi}
\def\text#1{%
\relax
\ifmmode %
\mathchoice
{\hbox{\everymath{\displaystyle}\rm #1}}%
{\hbox{\everymath{\textstyle}\rm #1}}%
{\hbox{\everymath{\scriptstyle}%
\def\prm{\fam\z@ \the\scriptfont\z@ \relax}%
\def\pit{\fam\itfam \the\scriptfont\itfam \relax}%
\rm #1}%
}%
{\hbox{\everymath{\scriptscriptstyle}%
\def\prm{\fam\z@ \the\scriptscriptfont\z@ \relax}%
\def\pit{\fam\itfam \the\scriptscriptfont\itfam \relax}%
\rm #1}%
}%
\else %
\leavevmode\hbox{#1}%
\fi
}
\def\bbox#1{%
\leavevmode\text{%
\textfont0 \the\textfont\bffam
\scriptfont0 \the\scriptfont\bffam
\scriptscriptfont0 \the\scriptscriptfont\bffam
\@temptokena\everymath \boldmath \everymath\@temptokena
{$\m@th\relax#1$}%
}%
}
\font\fivbf=cmbx5 \font\sixbf=cmbx6 \font\sevbf=cmbx7 \font\egtbf=cmbx8
\expandafter\def\expandafter\ixpt\expandafter{\ixpt
\scriptfont\bffam\sixbf \scriptscriptfont\bffam\fivbf}
\expandafter\def\expandafter\xpt\expandafter{\xpt
\scriptfont\bffam\sevbf \scriptscriptfont\bffam\fivbf}
\expandafter\def\expandafter\xipt\expandafter{\xipt
\scriptfont\bffam\egtbf \scriptscriptfont\bffam\sixbf}
\expandafter\def\expandafter\xiipt\expandafter{\xiipt
\scriptfont\bffam\egtbf \scriptscriptfont\bffam\sixbf}
\expandafter\def\expandafter\xivpt\expandafter{\xivpt
\scriptfont\bffam\tenbf \scriptscriptfont\bffam\sevbf}
\else %
\@revmess{message}{NFSS detected! Assuming NFSS.}
\def\text#1{%
\relax\ifmmode
\mathchoice
{\hbox{{\everymath{\displaystyle}#1}}}%
{\hbox{{\everymath{\textstyle}#1}}}%
{\hbox{{\everymath{\scriptstyle}\let\f@size\sf@size\selectfont#1}}}%
{\hbox{{\everymath{\scriptscriptstyle}\let\f@size\ssf@size\selectfont#1}}}%
\glb@settings
\else
\mbox{#1}%
\fi
}
\def\bbox#1{%
\relax\ifmmode
\mathchoice
{{\hbox{\boldmath$\displaystyle#1$}}}%
{{\hbox{\boldmath$\textstyle#1$}}}%
{{\hbox{\boldmath$\scriptstyle#1$}}}%
{{\hbox{\boldmath$\scriptscriptstyle#1$}}}%
\glb@settings
\else
\mbox{#1}%
\fi
}
\def\mathhexbox{\protect\mathhexbox@}
\def\mathhexbox@#1#2#3{\relax
\ifmmode\mathpalette{}{\m@th\mathchar"#1#2#3}%
\else\leavevmode\hbox{$\m@th\mathchar"#1#2#3$}\fi}
\fi
\def\bmit#1{\bbox{#1}}

\newdimen\@mydim \newdimen\@shiftdim
\newbox\@myboxa \newbox\@myboxb

\def\figcaption{\refstepcounter\@captype
  \@dblarg{\@figcaption\@captype}}
\long\def\@figcaption#1[#2]#3{\par\addcontentsline{\csname
  ext@#1\endcsname}{#1}{\protect\numberline{\csname
  the#1\endcsname}{\ignorespaces #2}}\begingroup
    \@parboxrestore
    \normalsize
  \@makefigcaption{\csname fnum@#1\endcsname}{\ignorespaces #3}
  \par\endgroup}

\long\def\@makefigcaption#1#2{
 \vskip 10pt
 \setbox\@myboxa\hbox{\small FIG. \thefigure.\kern 0.65em}
   \@mydim=\wd\@myboxa
 \setbox\@myboxb\hbox{\small #2} \advance\@mydim by \wd\@myboxb
\ifdim \@mydim > \hsize \advance\@mydim by -\wd\@myboxb
  \renewcommand\baselinestretch{1}\small\normalsize
{\parskip=0pt\noindent\unhbox\@myboxa\par}
\vskip-8.5pt
\noindent\@shiftdim=\hsize \advance\@shiftdim by -\@mydim
\hskip\@mydim\vbox{
\hsize=\@shiftdim{\noindent\unhbox\@myboxb}}
\else \hbox to \hsize{\hfil\box\@myboxa\box\@myboxb\hfil}
\fi\vskip-6pt}

\def\tblcaption{\refstepcounter\@captype
  \@dblarg{\@tblcaption\@captype}}
\long\def\@tblcaption#1[#2]#3{\par\addcontentsline{\csname
  ext@#1\endcsname}{#1}{\protect\numberline{\csname
  the#1\endcsname}{\ignorespaces #2}}\begingroup
    \@parboxrestore
    \normalsize
  \@maketblcaption{\csname fnum@#1\endcsname}{\ignorespaces #3}
  \par\endgroup}

\long\def\@maketblcaption#1#2{
 \vskip 10pt
 \setbox\@myboxa\hbox{TABLE \@Roman\c@table.\kern 0.65em}
\@mydim=\wd\@myboxa
 \setbox\@myboxb\hbox{#2} \advance\@mydim by \wd\@myboxb
\ifdim \@mydim > \hsize \advance\@mydim by -\wd\@myboxb
  \renewcommand\baselinestretch{1}\small\normalsize
\noindent\hangindent\@mydim\hangafter=1
\unhbox\@myboxa\unhbox\@myboxb\par
\else \hbox to \hsize{\hfil\box\@myboxa\box\@myboxb\hfil}
\fi}

\def\ps@headings{\let\@\mkboth\markboth
\def\@oddfoot{\rm 
\hfil \rm 
}}
\def\@evenfoot{{\rm 
\hfil \rm 
}}

\makeatother

\long\def\@makefntext#1{\parindent 0em
      \noindent \hbox to 0.3em{\hss$^{\@thefnmark}$}#1}

\begin{document}
\baselineskip = 13pt
\parskip0\baselineskip
\newskip\eqnskip
\eqnskip 0.5\baselineskip
\def\nB{{\bf B}\hspace{-0.24cm} /}
\def\idd{D\hspace{-0.23cm} /}
\def\ivv{v\hspace{-0.188cm} /}
\def\ipp{p\hspace{-0.188cm} /}
\def\iqq{q\hspace{-0.188cm} /}
\def\iep{\epsilon \hspace{-0.17cm} /}
\def\iva{\varepsilon \hspace{-0.175cm} /}
\def\dv{dv\hspace{-0.28cm}{\raisebox{1.5ex}{$\sim$}}}
\def\iz{Z\kern-0.45em Z}
\def\inn{I\kern-0.35em N}
\def\ip{I\kern-0.3em P}
\def\iq{I\kern-0.5em Q}
\def\ir{I\kern-0.35em R}
\def\xx{{\em x\kern-0.35em x}}
\def\ch{\raisebox{.4ex}{$\chi$}}
\def\larr{$\partial$ \hspace{-0.4cm}{\raisebox{1.5ex}{$\leftarrow$}}}
\def\ds{\displaystyle}
\def\skipback{\vskip -\parskip}
\newcommand{\stackunder}[2]{\mathrel{\mathop{#2}\limits_{#1}}}

\baselineskip=13pt
\thispagestyle{plain}

\begin{center}
{\large {\bf
Differential Renormalization of QCD in the \\
Background-Field Method$^\dagger $
}}
\end{center}
\vspace{3pt}

\begin{center}
Su-Long Nyeo\\
\vspace{-1pt}
{\small\em
Department of Physics, National Cheng Kung University,\\
\vspace{-2pt}
Tainan, Taiwan 701, R.O.C.
}\\
\end{center}

\begin{quotation}
\small
A short outline is given on the application of differential regularization 
to QCD in the background-field method. The derivation of the propagators in
the background gluon field as short-distance expansions is described and the
renormalization of the theory is mentioned.
\end{quotation}

\baselineskip=13pt

\footnotetext
{{\small
\hspace{-18pt} $^{\dag}$ Talk given at the $\,$JOINT $\,$JINR-ROC $\,
$(Taiwan) Conference on Intermediate and High Energy Physics, June 26-28, 
1995, Dubna, Russia.}
}

\vspace{8pt}

\noindent
{\normalsize \bf I. Introduction}
\vspace{12pt}

\hspace{0.3cm} Loop calculations in quantum field theories are generally 
plagued by infinities, and hence a regularization procedure is needed. The 
most commonly-used method is dimensional regularization, which respects 
gauge and Lorentz symmetries, but it is not suitable for chiral gauge 
theories like the electroweak theory or for calculations involving 
polarized particles.

\hspace{0.3cm} In this talk I will give a short outline on the application 
of a regularization method known as differential regularization (DR) [1] 
to QCD. This method was extensively studied in recent years and was 
demonstrated to be applicable to chiral theories. It enjoys the property 
that no counterterms are needed in the renormalization procedure, and 
hence the number of Feynman diagrams needed for a given loop calculation 
in DR is generally less than that in any other known regularization method.

\hspace{0.3cm} Here I will consider DR for QCD in the background-field 
method [2-7], which is a useful computational tool that allows one to 
compute radiative corrections while maintaining manifestly the symmetries 
of the theory under consideration. However, we should note that the 
calculational procedure in [1] using DR exhibits some inconveniences, 
such as that fixing the renormalization scale in gauge theories requires 
the implementation of Ward identities. I will present a calculational 
background-field approach with DR to study QCD, so as to eliminate the 
cumbersome step of implementing Ward identities in the calculations of 
Green functions. It is hoped that the background-field approach with 
DR could provide a more systematic method for calculating higher-loop 
orders in QCD. This talk is thus organized as follows. In section II, 
I will give a derivation of a short-distance expansion for the gluon 
propagator in the background gluon field, and then obtain the one-loop 
effective action. In section III, I will briefly mention the use of the 
method of differential regularization for loop calculations. Finally, 
a discussion section is also included.

\vspace{12pt}
\noindent
{\normalsize \bf II. Propagators in the background gluon field}
\vspace{12pt}

\hspace{0.3cm} Consider the action for the gluon field in Euclidean space

\def\theequation{1}
\begin{equation}
\begin{array}{l}
  {\bmit S}[A]=\ds \int d^4x\ds \, \frac{1}{4} \, F_{\mu \nu }^a\, 
F_{\mu \nu }^a ,
\\[1\eqnskip]
  F_{\mu \nu }^a\; = \partial _\mu A_\nu ^a - \partial _\nu A_\mu ^a + 
gf^{abc} A_\mu ^b A_\nu ^c\, .
\end{array}
\end{equation}
In the background-field method [2-7], we let $A_\mu ^a = B_\mu ^a 
+ Q_\mu ^a$ with $B_\mu ^a(x)$ being the background gluon field and 
$Q_\mu ^a(x)$ the quantum fluctuation gluon field. To quantize system 
(1), we use the background-field gauge $F^a[Q]= D_\mu ^{ab}[B]Q_\mu ^b(x)
=(\partial _\mu \delta ^{ab}-gf^{abc}B_\mu ^c(x))Q_\mu ^b(x)=0$ and 
consider adding to (1) the gauge-fixing term ${\bmit S}_{gf}[Q]=
\frac{1}{2} \int d^4xF^a[Q]F^a[Q]$, which corresponds to the Feynman 
gauge. Then we have

\def\theequation{2}
\begin{equation}
\begin{array}{ll}
  {\bmit S}[B+Q]+{\bmit S}_{gf}[Q]={\bmit S}[B]\hspace{-0.1cm} &+
\ds \int d^4x \left (\ds\frac{1}{2}(D_\mu ^{ab}Q_\nu ^b)^2+
gf^{abc}G_{\mu \nu }^aQ_\mu ^bQ_\nu ^c \right .
\\[2\eqnskip]
  &+gf^{abc}\left (D_\mu ^{ad}Q_\nu ^d\right )Q_\mu ^bQ_\nu ^c
\\[2\eqnskip]
  &\left . +\ds\frac{1}{4}\, g^2 f^{abc}f^{ade}
Q_\mu ^bQ_\nu ^cQ_\mu ^dQ_\nu ^e \right ) ,
\end{array}
\end{equation}
where $G_{\mu \nu }^a = \partial _\mu B_\nu ^a - 
\partial _\nu B_\mu ^a + gf^{abc}B_\mu ^bB_\nu ^c$ is the 
field strength for the background gluon field. The corresponding 
Faddeev-Popov ghost term reads

\def\theequation{3}
\begin{equation}
\begin{array}{ll}
  {\bmit S}_{gh}[B,Q,\bar{\eta },\eta ] &=
\ds \int d^4x\bar{\eta }^a(x)D_\mu ^{ac}[B]D_\mu ^{cb}[B+Q]\eta ^b(x)
\\[1.5\eqnskip]
  &=\ds \int d^4x\bar{\eta }^a(x)(D^2[B])^{ab}\eta ^b(x)
\\[1.5\eqnskip]
  &~~~- gf^{cbe}\ds \int d^4x\bar{\eta }^a(x)
D_\mu ^{ac}[B]Q_\mu ^e(x)\eta ^b(x) ,
\end{array}
\end{equation}
where $\bar{\eta }^a(x)$ and $\eta ^a(x)$ are the Faddeev-Popov 
ghosts. From (2), the gluon propagator in the background gluon field reads

\def\theequation{4}
\begin{equation}
\begin{array}{ll}
  \langle Q_\mu ^a(x)Q_\nu ^b(y)\rangle \hspace{-0.2cm} & =
\langle x\, |\left (\ds\frac{1}{P^2\delta _{\mu \nu }-
2G_{\mu \nu }}\right ) ^{ab} |\, y\rangle \equiv D_{\mu \nu }^{ab}(x,y) ,
\\[2\eqnskip]
  \quad \quad \quad \quad P_\mu ^{ab}&=-iD_\mu ^{ab}=
-i\left (\partial _\mu \delta ^{ab}+B_\mu ^{ab}\right ) ,
\end{array}
\end{equation}
where $G_{\mu \nu }^{ab}=gf^{acb}G_{\mu \nu }^c$ and $B_\mu ^{ab}=
gf^{acb}B_\mu ^c$; while from (3), the ghost propagator reads

\def\theequation{5}
\begin{equation}
  \langle \eta ^a(x)\bar{\eta }^b(y)\rangle = \langle x\, |\, 
\left(\ds\frac{1}{P^2} \right )^{ab} |\, y \rangle .
\end{equation}
For perturbative calculations, we need the gluon propagator (4), the ghost
propagator (5) and the Feynman rules for the vertices, which can be derived
from (2) and (3).

\hspace{0.3cm} Since we are interested in the ultra-violet divergences, 
we look for the short-distance expansions for propagators (4) and (5). 
We consider an expansion for the gluon propagator (4) in the following 
form ($z \equiv x - y$) [4]:

\def\theequation{6}
\begin{equation}
  D_{\mu \nu }^{ab}(x,y)=\ds\frac{1}{4\pi ^2}
\left (\ds\frac{1}{z^2}U_{\mu \nu }^{ab}(x,y)
  +V_{\mu \nu }^{ab}(x,y)\ln(M^2z^2)+W_{\mu \nu }^{ab}(x,y)\right ) ,
\end{equation}
so that it can be used in conjunction with DR method for loop calculations.
In (6), $U(x,y),~V(x,y)$ and $W(x,y)$ are analytic functions, which will be
expanded in series of $x$ and $y$, and the parameter $M$ plays the role of a
subtraction point.

\hspace{0.3cm} The gluon propagator satisfies

\def\theequation{7}
\begin{equation}
  \Delta _{\alpha \beta }^{ab}(x)D_{\beta \sigma }^{bc}(x,y)=
\delta ^4(x-y)\delta _{\alpha \sigma }\delta ^{ac} ,
\end{equation}
where

\def\theequation{8}
\begin{equation}
\begin{array}{ll}
  \Delta _{\alpha \beta }^{ab}(x) \hspace{-0.2cm} &=
\left [P^2\delta _{\alpha \beta }-2G_{\alpha \beta }\right ]^{ab}(x)
\\[1.5\eqnskip]
  &=-\Box \delta _{\alpha \beta }\delta ^{ab}-
\left [(\partial _\mu B_\mu (x))+2B_\mu (x)\partial _\mu +
B_\mu ^2(x)\right ]^{ab}\delta _{\alpha \beta }
  -2G_{\alpha \beta }^{ab}(x).
\end{array}
\end{equation}

\hspace{0.3cm} To determine $U(x,y),~V(x,y)$ and $W(x,y)$, we employ the 
Fock-Schwinger gauge $x_\mu B_\mu (x)=0$ to express $B_\mu (x)$ in terms 
of $G_{\mu \nu }(x)$. We can then rewrite (8), in matrix notation, as

\def\theequation{9}
\begin{equation}
  \Delta (x)=-\Box - X_\mu (x)\partial _\mu - Y(x) ,
\end{equation}
where $X_\mu (x)$ and $Y(x)$ are expanded about $x=0$ as follows:

\def\theequation{10}
\begin{equation}
\begin{array}{ll}
  X_\mu (x) = \hspace{-0.2cm} & x_\alpha G_{\alpha \mu }
  +\ds\frac{2}{3}x_\alpha x_\beta D_\alpha G_{\beta \mu }
  +\ds\frac{1}{4}x_\alpha x_\beta x_\rho D_\alpha D_\beta G_{\rho \mu }
\\[2\eqnskip]
  &+\ds\frac{1}{15}x_\alpha x_\beta x_\rho x_\sigma 
D_\alpha D_\beta D_\rho G_{\sigma \mu }
+{\cal O}(x^5) ,
\end{array}
\end{equation}

\def\theequation{11}
\begin{equation}
\begin{array}{ll}
  Y(x)=\hspace{-0.2cm} & 2G+2x_\alpha D_\alpha G+
x_\alpha x_\beta a_{\alpha \beta }+x_\alpha x_\beta x_\rho 
b_{\alpha \beta \rho }
\\[1\eqnskip]
  &+x_\alpha x_\beta x_\rho x_\sigma c_{\alpha \beta \rho \sigma }+
{\cal O}(x^5);
\end{array}
\end{equation}

\def\theequation{}
\begin{equation}
\begin{array}{ll}
  a_{\alpha \beta } \hspace{-0.1cm} & =
\ds\frac{1}{8}D_\gamma D_\alpha G_{\beta \gamma }+
\ds\frac{1}{4}G_{\alpha \gamma }G_{\beta \gamma }+D_\alpha D_\beta G,
\\[2\eqnskip]
  b_{\alpha \beta \rho } &=\ds\frac{1}{30}[D_\gamma D_\alpha +
D_\alpha D_\gamma ]D_\beta G_{\rho \gamma }+
\ds\frac{1}{6}[G_{\alpha \gamma }D_\rho +
(D_\rho G_{\alpha \gamma })]G_{\beta \gamma }
  +\ds\frac{1}{3}D_\alpha D_\beta D_\rho G ,
\\[2\eqnskip]
  c_{\alpha \beta \rho \sigma }&=
\ds\frac{1}{144}[D_\gamma D_\alpha D_\beta +D_\alpha D_\gamma D_\beta +
D_\alpha D_\beta D_\gamma ]D_\rho G_{\sigma \gamma }
  +\ds\frac{1}{9}(D_\rho G_{\alpha \gamma })(D_\sigma G_{\beta \gamma })
\\[2\eqnskip]
  &~~~+\ds\frac{1}{16}[G_{\alpha \gamma }D_\rho D_\sigma
  +(D_\rho D_\sigma G_{\alpha \gamma })]G_{\beta \gamma }
  +\ds\frac{1}{12}D_\alpha D_\beta D_\rho D_\sigma G .
\end{array}
\end{equation}
Here $G$ is a matrix with elements $gf^{acb}G_{\mu \nu }^c$. 
In obtaining (10) and (11), we
have used the equations of motion $D_\mu G_{\mu \nu }=0$.

\hspace{0.3cm} Solving (7), we get

\def\theequation{12}
\begin{equation}
  U(x,y) = I + \ds\frac{1}{2}x_\alpha y_\beta G_{\alpha \beta }(0) + 
\ds\frac{1}{6}x_\alpha(x_\beta + y_\beta )y_\rho D_\alpha 
G_{\beta \rho }(0) + \cdots ,
\end{equation}

\def\theequation{13}
\begin{equation}
\begin{array}{ll}
  V(x,y)=\hspace{-0.2cm} & -\ds\frac{1}{2}G(0) - \ds\frac{1}{4}
(x_\alpha + y_\alpha )D_\alpha G(0) + x_\alpha x_\beta V_{20\alpha \beta }
\\[2\eqnskip]
  &+y_\alpha y_\beta V_{02\alpha \beta } + x_\alpha y_\beta 
V_{11\alpha \beta } + \cdots ,
\end{array}
\end{equation}
where

\def\theequation{}
\begin{equation}
\begin{array}{lll}
  V_{20\alpha \beta } \hspace{-0.2cm} &=\hspace{-0.2cm} & V_{02\alpha \beta } = 
\ds\frac{1}{8}\delta _{\alpha \beta }G(0)G(0) ,
\\[2\eqnskip]
  V_{11\alpha \beta }&=&\ds\frac{1}{8}G(0)G(0)\delta _{\alpha \beta } + 
\ds\frac{1}{192}G_{\alpha \rho }(0)G_{\beta \rho }(0)
\\[2\eqnskip]
  &&-\ds\frac{\delta _{\alpha \beta }}{4}\left ( \ds\frac{3}{2}G(0)G(0) + 
\ds\frac{1}{16}G_{\rho \sigma }(0)G_{\rho \sigma }(0) + 
\ds\frac{1}{4}D_\rho D_\rho G(0) \right ) .
\end{array}
\end{equation}

\hspace{0.3cm} It can be seen that the function $W(x,y)$ is intimately 
related to $U(x,y)$
and $V(x,y).$ For example, by rescaling the parameter $M,~W(x,y)$ can change
by an amount proportional to $V(x,y)$. Here $W(x,y)$ is not needed.

\hspace{0.3cm} The short-distance expansion for the ghost propagator can be
obtained simply from the gluon propagator with $G(x)=0$.

\hspace{0.3cm} We can now use the short-distance expansions for the 
propagators to obtain the effective action up to one-loop order,

\def\theequation{14}
\begin{equation}
  \Gamma [B] \approx {\bmit S}[B] + \Gamma ^{(1)}[B] ,
\end{equation}
where $\Gamma ^{(1)}[B]$ is the one-loop effective action [3,5]

\def\theequation{15}
\begin{equation}
  \Gamma ^{(1)}[B] = \ds\frac{1}{2}\ln~{\rm Det}~D^{-1}_{\rm gluon}-
\ln ~{\rm Det}~D^{-1}_{\rm ghost} ,
\end{equation}
with $D^{-1}$ denoting the inverse of the propagator $D$. From the logarithmic
terms in the propagators, we obtain the derivative of the renormalized
one-loop effective action with respect to ln $M^2$:

\def\theequation{16}
\begin{equation}
\begin{array}{ll}
  \ds\frac{d\Gamma ^{(1)}[B]}{d\ln M^2}\hspace{-0.2cm} &=
\ds\frac{1}{2}~\ds\frac{11}{96\pi ^2} C_A
  \ds \int d^4xG_{\rho \sigma }^a (x) G_{\rho \sigma }^a (x)
  -\ds\frac{11}{96\pi ^2} C_A \int d^4 x
G_{\rho \sigma }^a (x)G_{\rho \sigma }^a(x)
\\[2\eqnskip]
  &=-\ds\frac{11}{48\pi ^2} C_A \ds\frac{1}{4}\int d^4 x
G_{\rho \sigma }^a(x)G_{\rho \sigma }^a(x) .
\end{array}
\end{equation}

\hspace{0.3cm} Finally, the one-loop $\beta $ function can 
be easily obtained by considering the
renormalization group equation for $\Gamma [B]$,

\def\theequation{17}
\begin{equation}
  M\ds\frac{\partial \Gamma [B]}{\partial M} + \beta (g) 
\ds\frac{\partial \Gamma [B]}{\partial g} = 0 ,
\end{equation}
which gives

\def\theequation{18}
\begin{equation}
  \beta (g) = -\ds\frac{11}{48\pi ^2} C_A g^3 .
\end{equation}

\hspace{0.3cm} Next, for simplicity, we briefly consider 
the massless quark Lagrangian density

\def\theequation{19}
\begin{equation}
  {\cal L} [ {\bar \psi }, \psi , A_\mu ^a ] = 
i\bar{\psi }\gamma _\mu D_\mu (A_\mu ^a)\psi .
\end{equation}
In the background-field method, we write

\def\theequation{20}
\begin{equation}
\begin{array}{ll}
  {\cal L}[\bar{\psi } + \bar{q},\psi  + q,B_\mu ^a + 
Q_\mu ^a ] =\hspace{-0.2cm} & {\cal L}[\bar{\psi },\psi ,B_\mu ^a] + 
i\bar{q}\gamma _\mu D_\mu (B_\mu ^a)q
\\[1\eqnskip]
  &+g(\bar{\psi }\gamma _\mu Q_\mu ^aT^aq + \bar{q}
\gamma _\mu Q_\mu ^aT^a \psi )
\\[1\eqnskip]
  &+g\bar{q} \gamma _\mu Q_\mu ^aT^a q ,
\end{array}
\end{equation}
where $\bar{\psi }$ and $\psi $ are the background quark fields, 
while $\bar{q}$ and $q$ are the quantum
fields. $D_\mu (B_\mu ^a) = \partial _\mu - igB_\mu ^aT^a$ is 
the covariant derivative in the background
gluon field $B_\mu ^a$ in the fundamental representation with 
$T^a$ being the generators.

\hspace{0.3cm} The massless quark propagator in the background 
gluon field is given by

\def\theequation{21}
\begin{equation}
  S(x,y) = \langle \, x\Bigl |\ds\frac{1}{\gamma _\mu P_\mu }\, 
\Bigr |y \rangle , P_\mu = -iD_\mu (B_\mu ^a) .
\end{equation}
The short-distance expansion for the quark propagator is more complicated
than that for the gluon's but can be found in many references [2-4,7]. I shall
refrain from listing its expression and proceed to the next section on the use
of DR method for higher-loop calculations or for operator renormalization.

\vspace{12pt}
\noindent
{\normalsize \bf III. The method of differential regularization}
\vspace{12pt}

\hspace{0.3cm} We saw that the one-loop effective action can be obtained by 
taking the
determinants of the propagators in the background field. However, calculations
of the effective action to higher-loop orders and of the evolution of operators
contain highly-singular terms of the form

\def\theequation{22}
\begin{equation}
  \ds\frac{1}{(z^2)^n}\ln^m (M^2 z^2), n\geq 2, m \geq 0,
\end{equation}
where $M$ is a mass parameter, which may be that in the expansions for the
propagators. The essential idea of DR method is to define the highly-singular
terms by

\def\theequation{23}
\begin{equation}
  \ds\frac{1}{(z^2)^n}\ln ^m(M^2z^2) = \underbrace{\Box \Box \cdots \Box }_{n-1} 
G(z^2), z^2 \neq 0,
\end{equation}
and to solve for the function $G(z^2)$, which has a well-defined 
Fourier transform
and depends on $2(n-1)$ integration constants.

\hspace{0.3cm} We list below the regularized expressions that are used in 
the loop
calculations:

\def\theequation{24}
\begin{equation}
  \ds\frac{1}{z^4} = -\ds\frac{1}{4}\Box \ds\frac{\ln (z^2M^2)}{z^2} ,
\end{equation}

\def\theequation{25}
\begin{equation}
  \ds\frac{1}{z^6} = \Box \Box H(z^2) ,
\end{equation}

\def\theequation{26}
\begin{equation}
\begin{array}{rl}
 H(z^2) \hspace{-0.2cm} & = -\ds\frac{1}{32}\ds\frac{\ln (z^2M_1^2)}{z^2} + 
M^2_2 \ln (z^2M_3^2) + M_4z^2 ,
\\[2\eqnskip]
  \ds\frac{1}{z^4} \ln (z^2M^2) & = \Box I(z^2) ,
\\[2\eqnskip]
  I(z^2) & = -\ds\frac{1}{8}\ds\frac{[\ln (z^2M^2)]^2 + 
2\ln (z^2M_1^2)}{z^2} + M_2 ,
\end{array}
\end{equation}
where $M_1, M_2, M_3$ and $M_4$ are arbitrary integration constants, 
which will be fixed by imposing gauge invariance. In the present 
background-field method, to fix a scale from the many integration 
constants, we only need to consider the gauge invariance of the 
background field, and not to use Ward identities as mentioned in [1]. 
The renormalized effective action at a given loop order is
then obtained by considering its derivative with respect to ln $M^2$ 
(cf. (16)).

\vspace{12pt}
\noindent
{\normalsize \bf VI. Discussion}
\vspace{12pt}

\hspace{0.3cm} In this talk I have briefly outlined the procedure of 
using differential regularization for loop calculations in the 
background-field method with the aim of providing a more systematic approach. 
It amounts to first writing the
propagators in the background gluon field as short-distance expansions in
Euclidean space. From the expansions, the one-loop effective action can be
immediately obtained. The propagators are expanded in such a way that they
can be used in conjunction with DR for loop calculations. We note that the
present approach enjoys an advantage of avoiding the use of Ward idcntities
for maintaining the gauge invariance of any loop calculations. The approach
is not limited to one-loop calculations, and the number of Feynman diagrams
needed for a given loop calculation in DR method is generally less than that
in any other regularization method, since DR needs no counterterms.

\hspace{0.3cm} We should note that for evaluating the effective action to 
one-loop order, there is no need to use DR, and the short-distance 
expansion procedure applies in either Euclidean or Minkowski space. Only 
when we calculate the two-loop or higher-loop orders that we need to 
regularize the singular terms,
and the use of DR requires working in Euclidean space.

\hspace{0.3cm} The approach mentioned in this talk can naturally be 
used for studying operator renormalization to higher-loop orders, for 
which one can easily extend the method in [2-4,6,7].

\newpage
\noindent
{\normalsize \bf Acknowledgments}
\vspace{12pt}

\hspace{0.3cm} It gives me great pleasure to thank Y.-H. Chen for 
discussions and for performing some computations. This work was 
supported by the National Science Council of the Republic of China 
under contract number NSC 84-2112-M006-013.

\vspace{12pt}
\noindent
{\normalsize \bf References}

\small
\begin{enumerate}
\item [{[$\;$1$\;$]}]
D. Z. Freedman, K. Johnson and J. I. Latorre, Nucl. Phys. {\bf B371}, 
353 (1992).
\vspace{-5pt}

\item [{[$\;$2$\;$]}]
E. V. Shuryak and A. I. Vainshtein, Nucl. Phys. {\bf B199}, 451 (1982); 
ibid.  {\bf B201}, 141 (1982).
\vspace{-5pt}

\item [{[$\;$3$\;$]}]
V. A. Novikov, M. A. Shifman, A. I. Vainshtein, and V. I. Zakharov, 
Fortschr.  Phys. {\bf 82}, 585 (1984); A. I. Vainshtein, V. I. 
Zakharov, V. A. Novikov, and M. A. Shifman, Yad. Fiz. {\bf 39}, 
124 (1984) [Sov. J. Nucl. Phys. {\bf 39}, 77 (1984)].
\vspace{-5pt}

\item [{[$\;$4$\;$]}]
W. Hubschmid and S. Mallik, Nucl. Phys. {\bf B207}, 29 (1982); J. Govaerts,
F. de Viron, D. Gusbin and J. Weyers, Nucl. Phys. {\bf B248}, 1 (1984).
\vspace{-5pt}

\item [{[$\;$5$\;$]}]
C. Lee, Nucl. Phys. {\bf B207}, 157 (1982); I. Jack and H. Osborn, Nucl.
Phys. {\bf B207}, 474 (1982).
\vspace{-5pt}

\item [{[$\;$6$\;$]}]
M. J. Booth, Phys. Rev. {\bf D4}2, 2518 (1992).
\vspace{-5pt}

\item [{[$\;$7$\;$]}]
I. I. Balitsky and V. M. Braun, Nucl. Phys. {\bf B311}, 541 (1988/89).
\end{enumerate}

\end{document}